\documentclass[12pt,a4paper]{article}
 \usepackage{nicefrac}
  \usepackage[margin=5mm,a5paper]{geometry}

\title{Dynamics of suspended rigid aggregating particles in flowing medium: theory, analysis and scientific computing}

\author{
Sarthok Sircar\thanks{Corresponding author \protect\url{mailto:sarthok.sircar@adelaide.edu.au}} 
\and
Anthony J. Roberts\thanks{School of Mathematical Sciences, University of Adelaide, South Australia~5005, Australia. 
\protect\url{mailto:anthony.roberts@adelaide.edu.au}}
}

\usepackage{microtype, amsmath, amssymb, url, subfigure}
\usepackage{graphicx,bm,empheq}
\usepackage[numbers]{natbib}
\usepackage[left]{lineno}

\renewcommand{\vec}[1]{\text{\boldmath$#1$}}
\newcommand{\ben}{\begin{equation}}
\newcommand{\een}{\end{equation}}

\begin{document}
\maketitle

\begin{abstract}
We develop and present a unified multi-scale model (involving three scales of spatial organisation) to study the dynamics of rigid aggregating particles suspended in a viscous fluid medium and subject to a steady poiseuille flow. At micro-level, the theory of adhesion describing the attachment / detachment kinetics of two rigid spheres coated with binding ligands, is utilized to describe the collision frequency function. The meso-scale dynamics is outlined through a continuous general dynamic equation governing the time rate of change of the particle size distribution function. The micro-meso coupling is achieved via the balancing of the mesoscale drag forces and couples with the micro-scale forces associated with the binder kinetics. Inside the macro domain (i.e., a long pipe), the model is equation free and divided into equal sized patches. The macroscale solution within each patch is obtained via appropriate (extrapolatory) coupling and amplitude conditions. 
\end{abstract}


\noindent {\bf Keywords:} surface adhesion, collision frequency function, Smoluchowski coagulation equations, multi-scale modelling, equation-free modelling, patch dynamics

\section{Introduction} \label{sec:intro}
The process of particle aggregation in the presence of fluid flow, is indispensible in many important scientific and industrial applications. The aerosol pollutant aggregation within the atmosphere~\citep{Friedlander1977}, cell aggregation in blood flow~\citep{Crowl2010}, rheology of liquid crystal polymer suspensions~\citep{Calderer2004,Forest2006,Sircar2009}, cluster growth of paramagnetic particles in microchannels~\citep{Promislow1995}, adherence of medical gels with nano-particles for targeted drug delivery~\citep{Keener2011a}, colloidal suspensions in pulp and paper-making industries as well as wastewater treatment plants~\citep{Somasundaran2005} are just some examples. The mechanism is spatio-temporally multi-scale, beginning with the coalescence of two surfaces at particle-scale (or micro-scale) and ensuing dynamics of particle agglomerates at the continuum-scale (or macro-scale). While the established experimental protocols have successfully described aggregation at macro-scale~\citep{Neelamegham1997}, a clear understanding of the multi-scale relationships between the microscopic phenomena of surface adhesion and the macroscopic structure of the aggregating flocs, is lacking~\citep{Marshall2007}. The complexities involved at each scale of spatial structure of the aggregates, necessitate the development of novel multi-scale theory that dynamically relays information between the fine(micro)-scale to the bulk(macro)-scale. Therefore the motivation of this article is to develop a single, unified theory and approach that can capture the complex, dynamical evolution of the process of aggregation at each scale of spatial organisation.

Past investigations in the micro-scale modelling of fluid-borne surface adhesion have addressed some theoretical challenges. These include the ligand-receptor binding kinetics~\citep{Dembo1988}, surface deformation~\citep{Hodges2002}, excluded volume effects~\citep{Poland1992}, paramagnetism~\citep{Promislow1995}, short range interactions~\citep{Zhang2003} and flow past the surrounding surfaces \cite{Goldman1967}. Consequently, many detailed kinetic models have successfully described the adhesion-fragmentation processes from the microscopic perspective. Schwarz~\citep{Korn2006} and more recently Mahadevan~\citep{Mani2012} studied the cellular adhesion between the ligand coated wall and a sphere moving in a shear flow. A similar model by Ranganathan~\citep{Ramesh2015} described surface adhesion via Langevin simulations. Although these micro-scale models correctly predict the fine-scale/particle level information, they have limited applications in validating large, industrial scale experiments~\citep{Somasundaran2005}.

On the meso-level of spatial organisation are computational methods that fully resolve the coupling between the particles themselves and between the particles and the fluid. One example is the direct numerical simulation of the momentum equations in which multiple rigid or elastically deformable particles are present~\citep{Hu2001}. Recently, lattice Boltzmann~\citep{Crowl2010} and dissipative particle dynamics~\citep{Pivkin2006} also have been employed for this purpose. While these direct approaches fully resolve the dynamics at the particle scale and accurately capture the hydrodynamic interactions between particles in an aggregate, they are computationally expensive to implement~\citep{Brown2008}. 

As a result, a slew of macro-scale, coarse-grained approaches have emerged that allow explicit aggregate morphology to be retained while removing the full coupling between individual particles and the fluid~\citep{Mousel2010, Marshall2007}. However, these approaches together with the continuum theories~\citep{Cogan2004}, rely on phenomenological assumptions governing the geometry of the aggregates in fluid flow, in place of rigorous upscaling of the micro-scale models, thereby limiting their predictive power outside the range of parameters for which these models have been calibrated. In summary, each of these research efforts have focused on the hierarchical spatial structure using separate theories. Efforts to link the coupled multiscale dynamics of the aggregation mechanism, are limited~\citep{Mori2013}. 


Sciortino made a recent effort to couple the microscopic ligand kinetics of charged surfaces with the meso-macro general dynamic equations (GDE) governing particle aggregation dynamics, but those numerical studies were done with chemically inert particles~\citep{Corezzi2012}. Other examples in this direction includes developing probabilistic extensions of the Smoluchowski's multiplicative aggregation kernel in one~\citep{Odriozola2001} and two dimensions~\citep{MonchoJorda2001}, with kernels containing one scaling parameter to be fit to data. Jia develop a method for predicting critical coagulant concentration via deriving a kernel incorporating surface charge density and potential as a function of the electrolyte~\citep{Jia2006}. Gilbert investigated and validated the forces and potentials for nanoparticles \cite{Gilbert2007} while Babler and Morbidelli studied aggregation and fragmentation, but only driven by diffusion and shear flow~\citep{Babler2007}. Using a single theory, the present work attempts to resolve the complications involving the aggregation process at three different scales of spatial organisation, i.e., the physical difficulty of modelling the surface adhesion of two particles at micro-level, the computational pitfall of modelling multi-particle aggregation at meso-level and the computational expense of simulating the resultant GDE in large macro-scale domain (i.e., a long pipe in the current study). An innovative technique to connect the spatio-temporal dynamics between the micro-meso spatial scales, via the balancing of the forces and torques, is described. The computational challenge, at macro-scale, is tackled by developing an `equation-free' patch dynamics method for the numerical macro-scale modelling of meso-scale system with fine scale details.


The article is laid out as follows. Section~\ref{sec:model} describes the comprehensive mathematical model. A physical model outlining particle collisions as function of particle size, is developed at micro-scale (\S\ref{subsec:micro}). The model explicitly accounts for the binding/unbinding of the tethers on the particle surface (\S\ref{subsubsec:Binder}) as well as the interaction of the charged surface with the ions dispersed in the fluid medium (\S\ref{subsubsec:LongRange}). An innovative mechanism between the micro-meso closure and the derivation of the collision frequency function is provided in \S\ref{subsubsec:Collision}. Next, the collision frequency function is then introduced into the GDE governing the spatially varying profile of particle size distribution function, \(n(u, {\vec x}, t)\) (\S\ref{subsec:meso}). A cross-sectionally averaged, slow manifold solution is derived in \S\ref{subsubsec:manifold}. Finally, a patch dynamics method to numerically evaluate the slow manifold solution in 1-D cross-sectionally averaged domain is discussed in \S\ref{subsec:macro}. The simulation results are highlighted in \S\ref{sec:results}, which is followed (in \S\ref{sec:conclusions}) with a brief discussion of the biophysical implication of these results and the focus of our future directions. 

\section{Mathematical model} \label{sec:model}
The aim in this study is to explore how the adhesion mechanism for two rigid, spherical particles governed by the micro-hydrodynamics as well as the microscale surface forces including the attractive/repulsive forces in an ionic medium, bond stretching and finite resistance to rotation via bond tilting, impact the population balance of the aggregates in mesoscale as well as the evolution dynamics in a cross-sectionally averaged pipe flow at macroscale. This aim is achieved by numerically calculating the particle size distribution function, \(n(u, {\vec x}, t)\). In the following discussion, the lowercase and uppercase quantities (e.g., {\vec x}, {\vec X}) will denote variables at the micro and the macro levels, respectively. The particle size distribution function is of our fundamental interest since most of the physical and chemical properties of flowing, particulate suspensions can be evaluated by taking appropriate moments, including the number concentration, mass concentration and aggregate structure tensors~\citep{Sircar2008}. 
\begin{figure}
\centering
\subfigure[Microscale]{\includegraphics[width=0.5\linewidth]{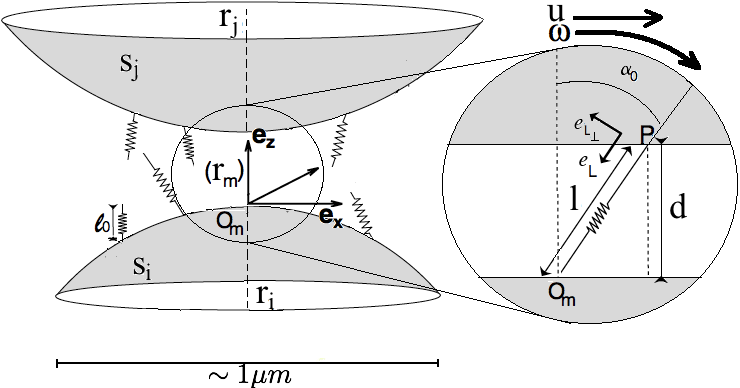}}
\subfigure[Mesoscale]{\includegraphics[width=0.25\linewidth]{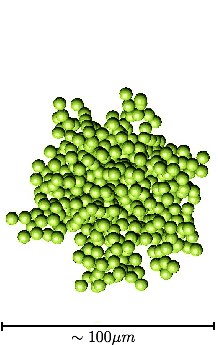}}
\subfigure[Macroscale]{\includegraphics[width=0.65\linewidth]{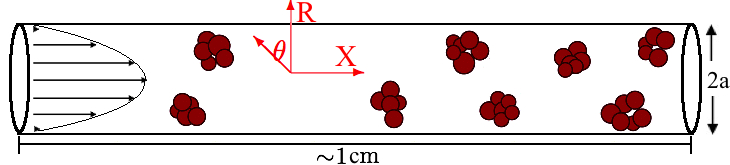}}
\caption{Illustration of aggregating particles moving in a flowing, charged medium, (a) Micro-level: Two spherical, rigid particles coated with binding ligands. The symbol~$(r_m)$ denotes the moving frame of reference, with the origin $O_m$ fixed on the surface of sphere~s$_i$ at the centre of the adhesion region, (b) Meso-level: dynamics of rigid aggregating particle clusters, and (c) Macro-level: evolution of the slow manifold of particle aggregates in a 3-D round pipe with steady poiseuille flow. Typical length scale of spatial resolution is underlined below.
Lowercase quantities (e.g. $x$) and uppercase quantities (e.g., $X$) denote variables at the micro and the macro levels, respectively.}\label{fig:Fig1}
\end{figure}

\subsection{Micro-scale kinetics} \label{subsec:micro}
In this section, we derive the collision frequency function that describe the particle-particle interactions, $\beta$ (eqn.~(\ref{eq:unsteadySmoluchowski}), \S\ref{subsec:meso}), using a combination of kinetic theory and problem-specific chemical, biological and physical sources of
interactions. Each particle constitutes a rigid spherical core onto which linear, hookean, spring-like binding ligands (i.e. binders) are attached and the surface of the coalescing particles are linked through these ligands. We neglect the shearing effect of the fluid flow on the mean rest length of the binders as well as the spatial dependence in the material parameters. The effects of gravity, non-specific forces acting on the particles, as well as the roughness of the particle surface are neglected. We assume that subject to a finite tilting, the binders are fixed on the particle surface.

The fluid medium as well as the surface of the particles are charged, but only the effects of Coulombic repulsion and Van der Waals attraction are incorporated (\S\ref{subsubsec:LongRange}). Other interactions including hydration effects, hydrophobic attraction, short range steric repulsion, and polymer bridging, which are absent in the length scales of our interest, are neglected~\citep{Gregory2006}. The binder kinetics is assumed to be independent of the salt concentration (i.e. the spring stiffness, $\lambda_0$ (\S\ref{subsubsec:Binder}) is independent of the charge-screening length, $\delta$, and the zeta potentials, $\psi_i$ (\S\ref{subsubsec:LongRange})). This implies that we are neglecting the electro-viscous stresses~\citep{Tabatabaei2010}. Finally, due to their relatively large micron-size scale, the binder kinetics of these particles are significantly different from the core-shell nano-crystal interactions, which are applicable at much smaller scales~\citep{Duval2008}.

\subsubsection{Binder kinetics} \label{subsubsec:Binder}
%
Figure~\ref{fig:Fig1}a illustrates the adhesion of two particles which are modelled as two rigid spheres, $s_i$ (with radius $r_i$) and $s_j$ (with radius $r_j$), respectively. To simplify the visualisation of the micro-level dynamics, consider a moving frame,~\(R_m\), with origin~\(O_m\) (sometimes referred as the \emph{Lagrangian} frame of reference) fixed on the surface of the sphere~\(s_i\) at a point equidistant from the edge of the separation gap. The unit vectors for this frame of reference are~\(\vec e_x\), \(\vec e_y\) and~\(\vec e_z\). 
For a given spatial point~\({\vec x} = (x, y, z)\) in this moving frame, the total relative velocity (of sphere \(s_i\) with respect to \(s_j\)) is \({\vec v}=u(\vec U)\vec e_x+r_i\omega(\vec U)\vec e_y\), where the relative translational velocity, \(u\), and the rotational velocity \(\omega\), are functions of the fluid velocity, \(\vec U\) (\S\ref{subsubsec:manifold}). 
Let~\(d(x)\) be the vertical separation gap between the two spheres. Define \(A_{\text{Tot}} g(x)\, d A\) as the number of bonds that are attached between the surfaces~\(A\) and \(A+dA\) at time~\(t\). \(A_{\text{Tot}}\)~is the total number of binding ligands. The function~\(g\) is synonymous with the term \emph{sticking probability} \citep{Somasundaran2005}. The total number of bonds formed is \(\int_{a_s} A_{\text{Tot}} g(x)\, d A\)\,, where~\(a_s=\pi r^2_s\) is the area of adhesion (refer \S\ref{subsubsec:Collision} for details on the radius of adhesion, \(r_s\)). In further description of the model we  denote \(d(x) \equiv d\), without loss of generalisation. 

The forward and reverse reaction rates for the binding ligands are then written as Boltzmann distributions, allowing highly stretched bonds to be readily broken by thermal energy fluctuations.
The kinetics are also influenced by the surface potential of the two charged surfaces.
Further, we cater for the ligands tilting by a finite angle~$\alpha_0$ with respect to the vertical direction.
This tilt is again expressed as a Boltzmann distribution,~$\mathcal{D}(\alpha_0)$, such that a bond may form between the two spheres for a given angle $\alpha_0 \in (-\frac{\pi}{2}, \frac{\pi}{2})$.
With these degrees of freedom, the bond attachment\slash detachment rates are 
\begin{align}
K_{\text{on}} (x) &= K_{\text{on,eq}} \exp \left[ \frac{- \lambda_s(l(x)-{l}_0)^2 + W(d)}{2{k}_B T} \right ]\mathcal{D}(\alpha_0), \nonumber \\
K_{\text{off}} (x) &= K_{\text{off,eq}} \exp \left[ \frac{ (\lambda_0 - \lambda_s)(l(x)-{l}_0)^2 + W(d) }{2{k}_B T} \right], \label{eqn:reaction_rates}
\end{align}
where ${k}_B$ is the Boltzmann constant, \(T\) is the temperature, $l_0$~is the mean rest length of the binders, $\lambda_0$~is the binder stiffness coefficient, and $\lambda_s$~is the spring constant of the transition state used to distinguish catch ($\lambda_0 < \lambda_s$) from slip ($\lambda_0 > \lambda_s$) bonds~\citep{Dembo1988}. $W(d)$~is the total surface potential described in \S\ref{subsubsec:LongRange}. As depicted in Figure~\ref{fig:Fig1}a, \(l = \sqrt{d^2 + x^2}\) is the length of a bond in a stretched configuration.
The energy associated with tilting a bond from its vertical position is~\((1/2)\lambda_{\theta} \alpha_0^2\), (\(\lambda_{\theta}\) being the torsion constant) and the corresponding Boltzmann distribution is 
\ben
\mathcal{D}(\alpha_0) = \exp\left(-\frac{\lambda_{\theta}\alpha_0^2}{2{k}_B T}\right) \frac{1}{D_0}\,, \quad \alpha_0 = \tan^{-1}\frac{x}{d}\,, \label{eq:tilt}
\een
where \(D_0 = \int^{{\pi}/{2}}_{-{\pi}/{2}} \exp\big[{-\frac{\lambda_{\theta}\alpha^2_0}{2{k}_B T}}\big]\, d \alpha_0 = \sqrt{\frac{2\pi k_B T}{\lambda_\theta}} \text{erf}\left(\frac{\pi}{2}\sqrt{\frac{\lambda_\theta}{2k_B T}}\right)\) is the normalization constant for all possible tilt orientations along the flow-direction. Ignoring non-equilibrium binding kinetics (i.e., $\frac{\partial g}{\partial t}=0$), in the limit of small binding affinity and abundant ligands on the binding surface (i.e., \(K_{\text{eq}}=\nicefrac{(A_{\text{Tot}} K_{\text{on,eq}})}{ K_{\text{off,eq}} } \ll 1\)), the evolution equation for the \emph{sticking probability} is~\citep{Dembo1988, Reboux2008},
\ben
{\vec v}\cdot \nabla g = A_{\text{Tot}} K_{\text{on}} - K_{\text{off}} g\,, \quad g = 0 \quad  \text{for }   x \ge r_s \,.
\label{eq:coll_fac}
\een
%

\subsubsection{Long range interactions} \label{subsubsec:LongRange}
%
For two charged spheres, of radii~\(r_i,r_j\), the potential due to the repulsive Coulombic forces in the gap of size~$d$ is~\citep{Gregory2006} 
\ben
W_{\text{C}}(d) = 2\pi \epsilon_0 \epsilon \psi^2_0 \Big(\frac{2r_i r_j}{r_i + r_j}\Big) e^{-\delta d}, \label{eq:CI}
\een
where $\delta$ is the Debye length, $\epsilon$ and~$\epsilon_0$ the dielectric constant of vacuum and the medium, respectively, and $\psi_0$, the average zeta potential of the diffuse cloud of charged counterions. The potential due to the Van der Waal forces for these spheres in the regime of close contact  is 
\ben
W_{\text{VW}}(d) = -\frac{A}{6d} \frac{r_i r_j}{r_i + r_j}, \label{eq:VWI}
\een  
where A is the Hamaker constant, measuring the Van der Waal `two-body' pair-interaction for macroscopic spherical objects. The total surface potential is \(W(d) = W_{\text{C}}(d) + W_{\text{VW}}(d)\), which is pair-wise attractive over short and long range, and pair-wise repulsive over intermediate range.
%

\subsubsection{Collision frequency function} \label{subsubsec:Collision}
First, the mesoscale hydrodynamic forces, \({\vec F}\), and torques, \({\vec T}\), on two rigid spheres in Stokes' flow are listed. These were derived by O'Neill~ \cite{ONeill1970} using lubrication theory. At leading order in \(\epsilon=\frac{d}{r_i} (\ll 1)\), these expressions are
%
%
\begin{eqnarray}
&{\vec F}^t =& -\frac{8\pi \mu r_i u}{5} \frac{\zeta(2+\zeta+2\zeta^2)}{(1+\zeta)^3} \ln(\epsilon) {\bf e}_x, \nonumber
\\
& {\vec T}^t =& - \frac{4\pi \mu r^2_i u}{5} \frac{\zeta(4+\zeta)}{(1+\zeta)^2} \ln(\epsilon) {\bf e}_y, \nonumber
\\
& {\vec F}^r =& - \frac{4\pi \mu r^2_i \omega}{5} \frac{\zeta(4+\zeta)}{(1+\zeta)^2} \ln(\epsilon) {\bf e}_x, \nonumber 
\\
& {\vec T}^r =& -\frac{16\pi \mu r^3_i \omega}{5} \frac{\zeta}{(1+\zeta)} \ln(\epsilon) {\bf e}_y, \nonumber
\\
& {\vec F}^f =& \frac{3\pi \mu r_i U(25\zeta+16)}{4\zeta} \Big(\frac{\zeta}{1+\zeta}\Big)^2  {\bf e}_x, \nonumber
\\
& {\vec T}^f =& 8\pi \mu r^2_i U\left(\frac{\zeta}{1+\zeta}\right)^3 \left(1 - \frac{3}{16} \left(\frac{\zeta}{1+\zeta}\right)^3 \right) {\bf e}_y, \label{eq:macroscale_F}
\end{eqnarray}
where $U$ is the cross sectionally averaged fluid flow (\S\ref{subsubsec:manifold}), \(\zeta = \frac{r_i}{r_j}\) and $\mu$ is the dynamic viscosity of the fluid. The superscripts \((t,r,f)\) denote components arising from relative translation, rotation and the flow of the surrounding fluid, respectively. These hydrodynamic forces must balance the forces arising from all the bond interactions,
%
%
\begin{eqnarray} 
{\vec F}_E(u,\omega,d) =& - \int_{a_s} g (1 - 1/l) \left[x\vec e_x + d\vec e_z \right] d A,\nonumber
\\
{\vec F}_C(u,\omega,d) =& \int_{a_s} g \frac{r_i r_j}{r_i + r_j} \left(-\frac{4\pi \epsilon_0 \epsilon \psi_0}{\delta} \exp^{-\delta d}+\frac{A}{6 {d}^2} \right) {\vec e}_z d A, \nonumber
\\
{\vec F}_T(u,\omega,d) =& (\nicefrac{\lambda_{\theta}}{(2\lambda_0{\it l}_0^2)}) \int_{{a_s}} ({g}/{l}^2) \alpha_0 \left[ -{d}{\vec e}_x + {x}{\vec e}_z\right] d A. \label{eq:microscale_F}
\end{eqnarray}
%
The subscripts \(i=E,C,T\) denote {\em extension}, {\em surface charges} and {\em torsion}, respectively. The adhesion area of the circular patch is given by \(a_s=\pi r^2_s\), where the adhesion radius, \(r_s\), is found by Mahadevan (details in supplementary material,~\citep{Mani2012}) using a scaling law argument of the `settling phase of the particles',
\ben
r_s = 2\left(\frac{{\it k}_B T {\it l}_0}{\lambda_0} \right)^{1/2} \left(\frac{1}{r_i}+\frac{1}{r_j}\right)(r_i+r_j)^{1/2}.
\een
The coupling across the scales is obtained via the global force balance on the spheres in the horizontal and vertical directions, and the torque balance about the center of mass of the spheres (i.e., forces and torques from eqn.~(\ref{eq:macroscale_F}) with forces from eqn.~(\ref{eq:microscale_F})), which solves for the unknown variables, \(u,\omega,d\), i.e.,
\begin{eqnarray} 
&&({\bf F}_E + {\bf F}_C + {\bf F}_T) \cdot {\bf e}_z = 0, \nonumber
\\
&&({\bf F}_E + {\bf F}_C + {\bf F}_T) \cdot {\bf e}_x + (F^r + F^t + F^s) = 0, \nonumber
\\
&&\frac{\beta r_i}{\zeta+1} ({\bf F}_E + {\bf F}_C + {\bf F}_T) \cdot {\bf e}_x + (T^r + T^t + T^s) = 0. \label{eqn:balance}
\end{eqnarray} 
Finally, in a Stokes regime, the collision frequency function, \(\beta\) (eqn.~(\ref{eq:unsteadySmoluchowski}) in \S\ref{subsec:meso}), is proportional to the total microscale force, \({\vec F}_{\text{Tot}}={\vec F}_E+{\vec F}_C+{\vec F}_T\), given by~\citep{Sircar2013}
\ben
\beta = \gamma_A \|{\vec F}_{\textrm{Tot}}\|, \label{eq:collisionFreq}
\een
where $\gamma_A=\frac{\lambda_0}{A_{\text{Tot}}K_{eq}A\pi\mu}$ is known as the \emph{aggregation contact efficiency parameter}~\citep{Somasundaran2005}. The operator \(\| \cdot \|\) denotes the magnitude of a vector.
We remark that in the case of static equilibrium (\(u=\omega=0\)) with no bond tilting effects on the binders (\({\lambda_\theta}^{-1}\rightarrow0\)) the evolution of \(g\) (eqn.~(\ref{eq:coll_fac})) trivially reduces to
\ben
g = \nicefrac{A_{\text{Tot}}K_{\text{on}}}{K_{\text{off}}} = K_{eq}\exp^{-\lambda_0(d-{\it l}_0)^2},
\een
and further, with perfect binding (\(l=d\rightarrow {\it l}_0\) so that \(g\rightarrow K_{eq}\)), the collision frequency reduces to
\begin{align}
\beta(\mathit{u},\mathit{v})_{d\rightarrow{\it l}_0}&=\gamma_A \|\sum_{i=E,C} F_i\|\pi r^2_s \nonumber
\\
&=\frac{4k_BT}{\mu}\frac{(\mathit{u}^{1/3}+\mathit{v}^{1/3})^2}{\mathit{u}^{1/3} \mathit{v}^{1/3}}\frac{1}{{\it l}_0A} \left(-\frac{4\pi \epsilon_0 \epsilon \psi^2_0}{\delta {\it l}_0} \exp^{-\delta {\it l}_0}+\frac{A}{6 {\it l}^2_0}\right),
\label{eq:kernal_long}
\end{align}
where $\mathit{u}=\frac{4\pi}{3}r^3_i$ and $\mathit{v}=\frac{4\pi}{3}r^3_j$ are the respective volumes of the spheres. In a neutral solution with no dissolved ions (i.e., the Debye length, $\delta^{-1} \rightarrow 0$), we have
\ben
\beta(\mathit{u},\mathit{v})_{d\rightarrow{\it l}_0, \delta^{-1}\rightarrow 0} = \beta^{Br} = \frac{2{\it k}_B T}{3\mu{\it l}^3_0}\frac{(\mathit{u}^{1/3}+\mathit{v}^{1/3})^2}{\mathit{u}^{1/3} \mathit{v}^{1/3}},
\een
which is the collision frequency function for Brownian diffusion for hard spheres commonly used in many theoretical and experimental studies of aggregation~\citep{Friedlander1977,Gregory2006}. To summarize, the micro-scale surface adhesion model of two particles has the following features:

\begin{itemize}
\item Each particle constitutes a rigid spherical core onto which linear, hookean, spring-like binding ligands are attached. The surface of the coalescing particles are linked through these ligands. The ligand kinetics is mediated by the bond formation/breakage rates and modelled using eqn.~\eqref{eq:coll_fac}.
\item The rigid core of the particles, is charged and suspended in an ionic medium. These effects are modelled via the repulsive Coulombic interactions (eqn.~\eqref{eq:CI}) and the attractive Van der Waal interactions (eqn.~\eqref{eq:VWI}).
\item The coupling between dynamics at the micro(particle)-level and the meso(cluster)-level, is achieved through the global balance of the meso-scale drag forces and couples (eqn.~\eqref{eq:macroscale_F}) with the micro-scale forces associated with the binder kinetics (eqn.~\eqref{eq:microscale_F}).
\end{itemize}

\subsection{Meso-scale dynamics} \label{subsec:meso}
Particle aggregation, diffusion together with convection determines the time rate of change of the particle size distribution function, \(n({\vec x}, t)\). In this section a continuous GDE, sometimes referred as the population balance equation~\citep{Smoluchowski1916}, is set up. By solving this equation for different initial and boundary conditions, the size distribution can be calculated for geometries and flow conditions of practical interest. 

For a continuous distribution, the collision rate between particles of size \(\mathit{u}\) and \(\mathit{v}\) depends on the collision frequency function, \(\beta(\mathit{u},\mathit{v})\), described in \S\ref{subsubsec:Collision}. The net rate of formation of particles of size \(\mathit{u}\), between the range \(\mathit{u}\)-\(\mathit{u}\)+d\(\mathit{u}\), is given by
\ben
\left[\frac{\partial n}{\partial t}\right]_{\textrm{agg}} = \phi(n) = \frac{1}{2}\int_{\mathit{u}_{\textrm{min}}}^{\mathit{u}-\mathit{u}_{\textrm{min}}}\hspace{-1cm} \beta(\mathit{u}-\mathit{v},\mathit{v})n(\mathit{u}-\mathit{v})n(\mathit{v})d\mathit{v} - n(\mathit{u}) \int_{\mathit{u}_{\textrm{min}}}^{\mathit{u}_{\textrm{max}}-\mathit{u}_{\textrm{min}}} \hspace{-1.5cm}\beta(\mathit{u},\mathit{v})n(\mathit{v})d\mathit{v}, \label{eq:steadySmoluchowski}
\een
where \(\mathit{u}_{\textrm{max}}\) and \(\mathit{u}_{\textrm{min}}\) are maximum and the minimum size of particle aggregates. The minimum size is trivially set to the size of 1 particle. The non-linear term, \(\phi(n)\) (eqn.~(\ref{eq:steadySmoluchowski})), represents the time rate of change of the size-distribution in the absence of convection or diffusion. Alternatively, they can be interpreted as the `steady-state' solution of the particle aggregates in a steady fluid flow (since \(\partial n/\partial t = U\partial n/\partial x\) where \(U\) is the uniform fluid speed and \(x\) is the distance in the direction of flow). The factor of \(1/2\) in eqn.~(\ref{eq:steadySmoluchowski}) accounts for the double counting of the collisions in the integral. The GDE for continuous meso-scale size distribution, with (constant coefficient, $D$) convective diffusion in an incompressible fluid with velocity \({\vec U(\vec x, t)}\), is
\ben
\frac{\partial n}{\partial t} +{\vec U}\nabla\cdot n= D\nabla^2 n + \phi(n)\label{eq:unsteadySmoluchowski}
\een
The GDE is a nonlinear, partial integro-differential equation coupled across multiple spatial scales. We remark that the numerical experiments are set-up in a regime far away from turbulent flows, where particle fragmentation due to split~\citep{Neelamegham1997} and erosion~\citep{Pandya1983} are non negligible. The solution to the cross-sectionally averaged slow manifold of the number density in steady pipe-flow, is provided in \S\ref{subsec:macro}.

%

\subsubsection{Evolution of a slow manifold} \label{subsubsec:manifold}
This section justifies and constructs a model of the long-term dispersion of the particle aggregates along a long pipe that is carried by Poiseuille flow down the pipe~\citep{Mercer1994}. This approach, supported by center manifold theory, can be readily extended to cater for reactive~\citep{Wright2012} or sedimenting material~\citep{Suslov1999} in long pipes of complex and varying geometry~\citep{Mercer1994}.

Consider the particle aggregates being carried by a fluid flow in a round pipe of radius \(a\)  (Figure~\ref{fig:Fig1}c). We assume that the particles are neutrally buoyant, i.e., the particles have no effect on the density of the fluid so that concentration differences do not cause density differences that could generate secondary flows. The along pipe flow is the Poiseuille flow, \(\vec U(R)=2U(1-R^2/a^2)\vec e_X\), where \(U\) is the cross-sectionally averaged downstream velocity (defined later). Using the pipe radius, \(a\), and the time associated with cross-pipe diffusion, \(\tau=a^2/D\), as the characteristic length and time scale, the particle conservation \textsc{pde}~(\ref{eq:unsteadySmoluchowski}) reduces to 
\begin{align}
\frac{\partial n}{\partial t} + U(R)\frac{\partial n}{\partial X} &= \frac{1}{R}\frac{\partial }{\partial R}\left(R\frac{\partial n}{\partial R}\right) + \frac{1}{R^2}\frac{\partial^2 n}{\partial \theta^2} + \frac{\partial^2 n}{\partial X^2}+\widetilde{\phi(n)} \nonumber \\
&=\mathcal{L}(n) + \frac{\partial^2 n}{\partial X^2}+\widetilde{\phi(n)},\label{eq:pouiseuille}
\end{align}
for pouiseuille flow. \(\widetilde{\phi(n)}=\frac{a^2}{D}\phi(n)\). The boundary condition is that of zero diffusive flux on the wall,
\ben
\frac{\partial n}{\partial R}=0 \quad \text{on}\quad R=1.
\een

Note that the equilibria, \(n^*\), of the \textsc{pde}~(\ref{eq:pouiseuille}) satisfies \(\widetilde{\phi(n^*)}=0\). We find a slow manifold about such an equilibria and hence deduce a model global in size distribution and local to small~\(\partial/\partial x \), corresponding to the low wavenumber solution in Fourier space. Choosing to write the model in terms of the cross-pipe average, to an initial approximation, the slow manifold model is the trivial
\ben	
n^0\approx \overline{n}(X,t)\quad\text{such that}\quad \frac{\partial \overline{n}}{\partial t} \approx \widetilde{\phi(\overline{n}}),
	\label{svdispcm0}
\een
where the amplitude \(\overline{n}\) is defined via the projection, \(\overline{n}(X,t)\!=\!\!\int_0^1\!\!\!\int_0^{2\pi}\!\!\!n(R,\!\theta,\!X,\!t)\!R d\theta dR\)\!, which is the cross-pipe average. Now seek to iteratively refine this model~\eqref{svdispcm0}, using the corrections \({\hat n}(N,R,\theta)$~and~${\hat g}(N)\), where
\ben
 	n^1\approx n^0+{\hat n}
     \quad\text{such that}\quad 
     \frac{\partial \overline{n}}{\partial t}\approx \widetilde{\phi}+{\hat g}(\overline{n})\,, \label{eq:step1}
\een
is a more refined model. Substituting into the advection diffusion \textsc{pde}~(\ref{eq:pouiseuille}) and neglecting products of the small corrections (i.e., \(-U\partial \hat{n}/\partial X, \partial^2 \hat{n}/\partial X^2\)) gives
\ben
\frac{\partial n^1}{\partial t} = \mathcal{L}(n^1) - U(r)\overline{n}_X + \overline{n}_{XX} + \widetilde{\phi}, \label{eq:step2}
\een
where the subscript, $X$, denotes the longitudinal derivatives (see Figure~\ref{fig:Fig1}c). Using the approximation~(\ref{eq:step1}), the chain rule, \({\partial n^1}/{\partial t}=\left({\partial n^1}/{\partial N}\right)\left({\partial N}/{\partial t}\right)\approx\widetilde{\phi}+\hat{g}\) and the identity \(\mathcal{L}(\overline{n})=0\), we have 
\ben
\mathcal{L}(\hat{n})-\hat{g} = U(r)\overline{n}_X - \overline{n}_{XX}. \label{eq:step3}
\een
Note that \(\mathcal{L}\)~is self-adjoint, upon defining the inner product $\langle v,w\rangle=\int_0^1\int_0^{2\pi} vw\, Rd\theta\,dR$. An adjoint eigenvector of this operator is $z(R,\theta)\!=$constant.  Projecting eqn.~(\ref{eq:step3}) in the space spanned by this eigenvector, gives
\begin{displaymath}
-\pi {\hat g}=\overline{n}_X\int\!\!\!\int U(R)\,R\,d\theta\,dR -\pi \overline{n}_{XX}\,,
\end{displaymath}
which simplifies to
\ben
 {\hat g}=-U\overline{n}_X+\overline{n}_{XX}\,,
\een
as $U(R)=2U(1-R^2)$ has cross-sectional average of~$U$ by the non dimensionalisation. Using this update~${\hat g}$ and the fact that there are no $\theta$ variations in the right-hand side of eqn.~(\ref{eq:step3}), the first correction~${\hat n}$ can be found as ${\hat n}=w_1(R)U\overline{n}_X$ where $w_1(R)=\frac{1}{24}\left(-2+6R^2-3R^4\right)$. The arbitrary constant of integration is found from the constraint that the cross-sectional average of every correction~${\hat n}$ has to be zero. Correcting the initial model~\eqref{svdispcm0} with these ${\hat n}$~and~${\hat g}$ we deduce a refined model
\begin{equation}
	n^1\approx \overline{n}+w_1(R)U\overline{n}_X
	\quad\text{such that} \quad
	\frac{\partial \overline{n}}{\partial t}\approx -U\overline{n}_X+\overline{n}_{XX}+{\tilde \phi}\,.
	\label{svdispcm1}
\end{equation}
The Approximation Theorem suggests that the $N_{XX}$~term needs correction because the residuals in the equation resulting from this approximate model, (see the iterate, eqn.~(\ref{svdispcm2})), are of order 2~\citep{Mercer1994}. Hence, following the above procedure, a second level of refinement leads to
\begin{subequations}
\label{eq:reducedModel}
\begin{empheq}[]{align}
	n\approx \overline{n}+w_1(R)U\overline{n}_X+w_2(R)U^2\overline{n}_{XX} \label{eq:fastManifold} \\
	\frac{\partial \overline{n}}{\partial t}\approx -U\overline{n}_X+\left(1+\frac{U^2}{48}\right)\overline{n}_{XX}+{\widetilde \phi} \label{svdispcm2}
\end{empheq}
\end{subequations}
where $w_2(R)=\frac{1}{11520}\left(31-180R^2+300R^4-200R^6+45R^8\right)$. Higher order corrections of the estimate in eqn.~(\ref{svdispcm2}) can be easily derived using computer algebra.\footnote{\url{http://reduce-algebra.com/}} The reaction diffusion eqn.~(\ref{svdispcm2}) has an effective diffusion coefficient of $1+U^2/{48}$, i.e., the effective rate of dispersion along the pipe is faster than what is predicted by simple cross-sectional averaging. The dimensional form of the model~\eqref{svdispcm2}, is
\begin{equation}
 \frac{\partial \overline{n}}{\partial t}\approx-U\overline{n}_X+\left(D+\frac{U^2a^2}{48D}\right)\overline{n}_{XX}+\phi(\overline{n})\,.
\label{eq:dispfpm}
\end{equation}
The extra dispersion (with coefficient $U^2a^2/(48D)$), also called the shear dispersion, is proportional to the square of the velocity and inversely proportional to the cross-pipe diffusivity and implies  that the smaller the molecular diffusion, the larger the shear dispersion, a paradox which is well established~\cite{Roberts2003}.

\subsection{Macro-scale evolution: patch dynamics} \label{subsec:macro}

\begin{figure}[htbp]
\centering
\subfigure[Mesodomain with multiple patches]{\includegraphics[width=0.5\linewidth]{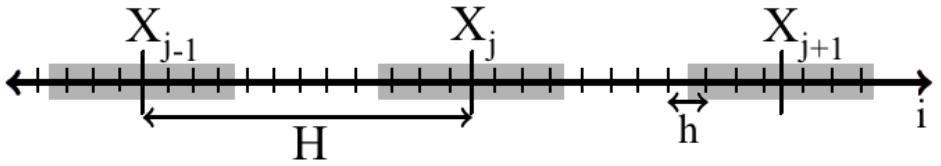}}
\subfigure[Closeup of one patch]{\includegraphics[width=0.5\linewidth]{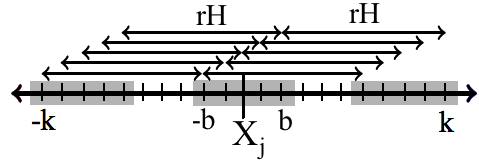}}
\caption{(a) Mesodomain with multiple stationary and equidistant patches. The meso-scale lattice is indexed by $\mathit{i}$ and indicated with short ticks on the horizontal axis with spacing $h$. The macro-scale lattice is indexed by $X_j$ and indicated with long ticks on the horizontal axis with spacing $H$.~(b) Close of one patch, with patch half-width $k=10$ and buffer half-width $b=2$. The shaded region in the centre ($i=0,\pm 1,\ldots,\pm b$) is the patch core and required for condition~(\ref{eq:amplitude}). The two outlined regions on each ends of the patch ($i=\pm(k-2b),\ldots,\pm k$) are the buffers and required for condition~(\ref{eq:couplingConditions}). Each point in each buffer is at a distance of $rH = (k-b)h$ from one point in the core.}\label{fig:Fig2}
\end{figure}

\subsubsection{Solution within each patch} \label{subsubsec:patch1}
We construct $j$th patch of width $(2k+1)h$ and denote it with macro-scale lattice point $X_j$ (Figure~\ref{fig:Fig2}a). $k$ is the patch half-width and $h$ is the meso-scale lattice spacing. First step, for each point within a patch and for every time-step, the non-linear aggregation term, $\phi(n)$ (eqn.~(\ref{eq:steadySmoluchowski})), is calculated. The integrals 
(eqn.~(\ref{eq:microscale_F})) in the collision frequency function (eqn.~(\ref{eq:collisionFreq})) using the adaptive Lobatto quadrature (via Matlab function {\it quadl}). The function, $\phi(n)$, is numerically evaluated by employing the Galerkin approximation scheme developed by Banks~\citep{Kappel1989} and adopted by others~\citep{Sircar2010}. While evaluating the integrals in eqn.~(\ref{eq:steadySmoluchowski}), we chose $\mathit{u}_{\text{min}}=1$ femtoliter (fL) as a lower bound in our simulations, since the radius of the smallest aggregate (consisting of a single particle) was of the order of a micron. In our discrete aggregation model we chose the upper bound, $\mathit{u}_{\text{max}}=2000$ fL and the volume resolution of $N=100$ bins per fL. Other parameters used in the simulation are listed in Table~\ref{tab:parameter}~(see \S\ref{sec:results}). As reported in~\citep{Kappel1989}, the convergence of this first order approximation scheme was tested and a linear (first order) relationship was found between the $l^\infty$-error of the solution and the mesh-size, $\delta u$ ($\delta u$ being the discretisation in the aggregate volume). The non-dimensional advection-reaction-diffusion equation~(\ref{svdispcm2}) is discretised using the standard Crank-Nicolson scheme. The non-linear reaction term is treated explicitly. Although the size of the domain is arbitrarily fixed at $K=100$ patches, the reason for the choice the patch half-width, $k$, buffer half-width, $b$, and the lattice spacings, $h, H$, is explained in \S\ref{sec:results}. Physical inlet and exit conditions are needed for the macro-scale solution in the first patch, $X_1$, and the last patch, $X_K$, respectively. The conditions are~\citep{Roberts1992},
\begin{subequations}
\label{eq:BC}
\begin{empheq}[]{align}
  N_1(t) &\approx  \frac{1}{\pi \kappa}e^{-\frac{t}{\kappa}}\!\ast\!\overline{U(r)n_1(t)} + \frac{1}{\pi \kappa}\left[\overline{U(r)r_1n_1(t)}-\frac{1}{\kappa}e^{-\frac{t}{\kappa}}\!\ast\!\overline{U(r)r_1n_1(t)}\right] \label{eq:inlet} \\
\frac{\partial N}{\partial X} &\approx e^{-\frac{t}{\kappa}} \ast \frac{\partial N}{\partial t}, \qquad \text{at}~ j=K \label{eq:outlet}  
\end{empheq}
\end{subequations}
where $\kappa=2/105$, $n_1(t)$ is the prescribed particle size distribution across the domain entrance. The overbar (\,$\overline{\phantom{w}}$\,), denotes the cross-stream average and the operator $\ast$ denotes the convolution, $f(t)\ast g(t) = \int_0^t f(\tau)g(t-\tau) d\tau$. The functions $r_1$ is  given by
\ben
r_1(R) = \frac{1}{280}\left(35 R^4 - 70 R^2 + 11\right), \label{eq:r1}
\een
The inlet condition~(\ref{eq:inlet}) ensures that the second order shear dispersion model~(\ref{eq:reducedModel}) is accurate after the entry transients have disappeared and the exit condition~(\ref{eq:outlet}) ensures that the exit does not have an upstream influence (via diffusion)~\citep{Roberts1992}. In following discussions, we drop the overbar over the variables, so that \(n=n(x,t)\) is the cross-sectionally averaged particle size distribution function.

\subsubsection{Extracting macroscale field} \label{subsubsec:patch2}
After eqn.~(\ref{eq:dispfpm}) is solved within the $j$th patch for $n_{ij}$, together with the coupling conditions (\S\ref{subsubsec:patch3}), we define a macro-scale solution field within each of these patches. We propose a macro-scale solution, referred as the patch amplitude from hereon, as an average over the meso-scale field within the patch `core' (Figure~\ref{fig:Fig2}b),
\ben
N_j(t) = N(X_j, t) = \sum^{i=b}_{i=-b}\frac{n_{ij}}{2b+1}, \label{eq:amplitude}
\een
where $b$ is the patch-buffer half width, such that $0\le b <\,$k. Some patch dynamics methods extract the macro-scale solution by trivially using the meso-scale solution at the patch center, corresponding to $b = 0$~\citep{Roberts2001}. However, unlike the `rough' micro-scale structure in the present setup, these methods are for `smooth' micro-scale dynamics with only slight variations within patches.

\subsubsection{Coupling condition between patches} \label{subsubsec:patch3}
Two coupling conditions are required to solve eqn.~(\ref{eq:dispfpm}), at each end within the $j$th patch. These conditions are derived from an interpolation of the macro-scale field, $N_j$, from its nearest macro-scale neighbours, $N_{j\pm1}, N_{j\pm2}, \ldots$ To write these conditions, we define a fractional step $r$, connecting the spacing between the meso-scale domain, h, and the macro-scale domain, H, such that $rH = (k-b)h$ (Figure~\ref{fig:Fig2}b). We also define the meso-scale and the macro-scale step operators, $\varepsilon, \overline{\varepsilon}$, such that $\varepsilon^{\pm 1}n_{ij}=n_{i\pm 1 j}$ and $\overline{\varepsilon}^{\pm 1}N_j=N_{j\pm 1}$, respectively. Thus, $\varepsilon^{\pm(k-b)}N_j=\overline{\varepsilon}^{\pm r}N_j$. Finally, we introduce the macro-scale mean and difference operators, $\overline{\mu}=(\overline{\varepsilon}^{1/2}+\overline{\varepsilon}^{-1/2})/2$ and $\overline{\delta}=(\overline{\varepsilon}^{1/2}-\overline{\varepsilon}^{-1/2})/2$, respectively. Using the condition~(\ref{eq:amplitude}), note that
\begin{align}
\varepsilon^{\pm(k-b)}N_j &=\sum^{i=k}_{i=k-2b}\frac{n_{\pm ij}}{2b+1} =\overline{\varepsilon}^{\pm r}N_j \nonumber \\
&= (\overline{\mu}\overline{\delta}+\frac{1}{2}\overline{\mu}^2+1)^{\pm r}N_j \nonumber \\
&\approx N_j + \sum_{q=1}^\Gamma \prod_{l=0}^{q-1}(r^2-l^2) \frac{\pm(2q/r)\overline{\mu}\overline{\delta}^{2q-1}+\overline{\delta}^{2q}}{(2q)!}N_j + \mathcal{O}({\Gamma+1}), \label{eq:couplingConditions}
\end{align}
which are the required coupling conditions.
The $\approx$ can be replaced with an equality provided $\Gamma\rightarrow\infty$. 
Detailed discussion on the theoretical support for patch dynamics are provided elsewhere~\citep{Roberts2001}. Finally, it has been shown that it is possible define arbitrary patch coupling conditions, provided suitably large buffers are chosen to shield the patch `core' from unwanted consequences of these coupling conditions over short evaluation times~\citep{Samaey2006}. To summarise, given the macro-scale solution, $N_j(t_{n-1})$ at time $t_{n-1}$, the patch dynamics algorithm for finding the macroscale solution , $N_j(t_n)$ at time $t_n$, is
\begin{enumerate}
\item Using the micro-scale field variables evaluated at time $t_{n-1}$, setup the non-linear reaction term, $\phi(n)$ (eqn.~(\ref{eq:steadySmoluchowski})), within each patch $j$. 
\item Numerically solve eqn.~(\ref{eq:dispfpm}), together with coupling conditions~(\ref{eq:couplingConditions}) within all patches $j$, such that $N_j=N_j(t_{n-1})$.
\item From the amplitude condition~(\ref{eq:amplitude}), determine the macro-scale solution at time $t_n$ for all macro-scale points $X_j$ (Figure~\ref{fig:Fig2}a), such that $n_{ij}=n_{ij}(t_n), N_j=N_j(t_n)$.
\item Using the macro-scale solution $N_j(t_n)$ evaluated in Step~3, return to Step~1 to determine the solution at time step $t_{n+1}$.
\end{enumerate}
The computational technique is `equation-free' since no attempt is made to derive an explicit macro-scale model. The procedure simply evaluates `on the fly' numerical solutions of the model at each macro-scale point, $X_j$, and time-step, $t_n$.

\section{Numerical results: dynamics of cross~sectionally averaged size distribution} \label{sec:results}
Table~\ref{tab:parameter} lists the parameters used in our numerical calculations.
The parameter values are chosen so that they closely replicate the adhesion-fragmentation of neutrophiles in slow viscous flow conditions. 
For example, the \textsc{p}-selectine molecule extends about \(40\)\,nm from the endothelial cell membrane, so when combined with its ligand \textsc{psgl}-1 it is reasonable to take ${l}_0 \approx 100$\,nm as an estimate of the length of the unstressed bond~\citep{Shao1998}.
Hochmuth measured variations of up to three orders of magnitude \emph{in vivo} in measuring the values of the  microvillus stiffness,~$\lambda_0$~\citep{Shao1998}.
Direct measurements of the parameters,~$A_{\text{Tot}}$, $K_{\text{on,eq}}$ and~$K_{\text{off,eq}}$ are scarce, although values in several thousands have been used in previous models~\citep{Hammer1996}.
Since we do not wish to study the effects of finite rotation of the ligands~\citep{Reboux2008} or the effect of catch-versus-slip bonds~\citep{Dembo1988}, the corresponding parameters related to these material properties are fixed at $\tilde{\lambda_{\theta}}=1.0$ and $\tilde{\lambda_s}=0.5$\,, respectively.
The dielectric constant in vacuum is $\varepsilon^D_0 = 8.854 \times 10^{-12}$\,Farad\,m$^{-1}$, whereas the permittivity of water at temperature~$25^\circ$C is $\varepsilon^D=78.5$ (not to be confused with~$\epsilon$ which is a length ratio defined in \S\ref{subsubsec:Collision} (eqn.~(\ref{eq:macroscale_F})) or the macroscale step operator, $\overline{\varepsilon}$, defined in \S\ref{subsubsec:patch3}).
The dissolved salt (furnishing the ions in the fluid) is assumed to be a 1-1 electrolyte with a zeta potential of $\psi_0=25$\,mV (corresponding to the surface potential studies by Gregory~\citep[Chap.~3]{Gregory2006}).
We assume that the solute concentration in the fluid only effects the Debye length,~$\delta$.
The Boltzmann factor is taken as ${k}_BT=3.1\times 10^{-21}$\,J. Finally the cross-sectionally averaged fluid speed is fixed at $U=1.5$ (eqn.(~\ref{svdispcm2})), to ensure the correct application of the inlet and exit conditions on the macro-domain (eqn.(~\ref{eq:BC})), developed by Roberts~\citep{Roberts1992}.
\begin{table}
\centering
\begin{tabular}{|c|c|c|c|c|}
\hline
Parameter & Value & Units & Source\\
\hline
A & 2.44 {\it k}$_B$T & J & \cite{Gregory2006} \\
A$_{\text{Tot}}$ & $10^9$ & $\text{m}^{-2}$ &\citep{Hammer1996} \\
K$_{\text{on, eq}}$ & $10^2$ & $\text{s}^{-1}$ &\citep{Hammer1996} \\
K$_{\text{off, eq}}$ & $10^{14}$ & $\text{s}^{-1}$ &\citep{Hammer1996} \\
$\lambda_0$ & $10^{-5}$--$10^{-2}$ & $\text{N\,m}^{-1}$ &\citep{Mani2012} \\
$\mu$ & $10^{-3}$ & $\text{N\,s\,m}^{-2}$ &\citep{Reboux2008} \\
$l_0$ & $10^{-7}$ & $\text{m}$ &\citep{Shao1998} \\
$U$ & 1.5 & -- &\citep{Roberts1992} \\
\hline
\end{tabular}
\caption{Parameters common to all numerical results. The first 7 parameters are used in the system of eqns.~(\ref{eq:Maineqn}--\ref{eq:collisionFreq}), while the last parameters is utilized in the time-dependent \textsc{pde}~(\ref{svdispcm2}).}\label{tab:parameter}
\end{table}

\subsection{Case $\phi(n)=0$: error analysis} \label{subsec:errorAnalysis}
\begin{figure}
\centering
\includegraphics[width=0.65\linewidth]{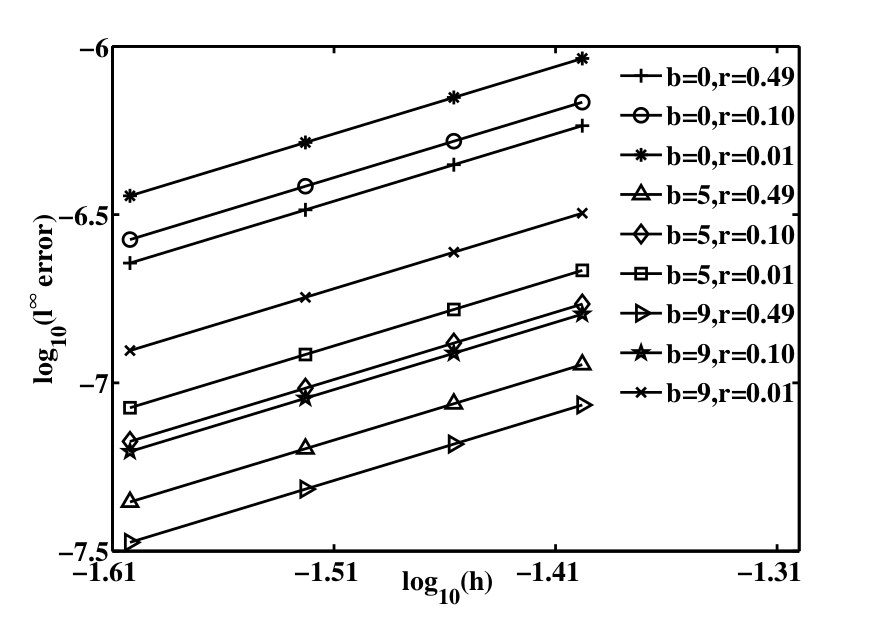}
\caption{ $l^\infty$ error versus meso-scale spacing, $h$, on $\log$-scale for different choices of patch buffer-half width, \(b=0, 5, 9\), and the ratio relating the meso-scale with the macro-scale spacing, \(r=0.49, 0.10, 0.01\). All curves have slope=2, indicating second order accuracy of the numerical method.}\label{fig:Fig3}
\end{figure}

First, the overall performance of the numerical method is tested for the advection-diffusion system, since an analytical solution is available in this case. We choose an initial profile, $n(x=0,t=0)=1$, the numerical time-step, $\triangle t=10^{-4}$. The $l^\infty$ error is evaluated for four different values of the meso-scale spacing, $h=0.025, 0.03, 0.035, 0.04$, which removes the possibility of any spurious oscillations of the approximate solution, as predicted by Von Neumann stability analysis (i.e., $\frac{D_0\triangle t}{h^2}<0.5$, where $D_0=1+\frac{U^2}{48}$, eqn.~(\ref{svdispcm2})). The application of Crank-Nicolson method implies second order accuracy of the method within each patch. Figure~\ref{fig:Fig3} highlights that the method is second order accurate in the macro-scale, as well (since all curves in the $\log-$scale of the $l^\infty$ error versus meso-scale spacing, $h$, has slope=2). The patch half-width is fixed at $k=10$. 
To choose an optimal value of the buffer half-width, $b$, and the meso-scale macro-scale spacing ratio, $r$, notice that since the patches cannot overlap, $0<r<0.5$. Further, lower values of $r$ implies that the patches are further apart. Lower value of $r$ (i.e., $r=0.01$) and no averaging within each patch (i.e., $b=0$, the curve ($-\ast-$) in Figure~\ref{fig:Fig3}) implies that the information is relayed between the scales with minimum computational effort. Consequently the numerical error is highest in this case, among all selected choices of ($b, r$). Alternatively, close patches (i.e., $r=0.49$) with maximal buffer half-width size (i.e., $b=9$, the curve ($-\triangleright-$)) leads to the lowest error with the selected values of ($b, r$), but the advantage of patch scheme (to perform computations widely spaced patches, is lost). We choose $b=5, r=0.10$, (the curve ($-\Diamond-$)) which keeps the numerical error reasonably low while maintaing the advantage of the method. Our choice of $b \approx k/2$ has been corroborated as an ideal one, in other applications~\citep{Roberts2001,Samaey2006}.

%

\section{Conclusion} \label{sec:conclusions}
By analysing the emergent, large scale, evolution we showed that patch dynamics macroscale modelling is able to capture the emergent dynamics of a microscale lattice system when the microscale  has fine detail. For best results it is important to appropriately choose the patch geometry, defined by the patch half-width $n$, and the patch coupling conditions, defined by the buffer half-width $b$, relative to the underlying period $K$ of the microscale detail. We showed that the symmetry of the microscale model is important and should be reflected in the choice of patch geometry and coupling conditions. We expect that other microscale models require similar consideration when implementing patch dynamics for macroscale modelling.


\paragraph{Acknowledgements} SS acknowledges the financial support from Dr. David Bortz in Dept. of Applied Mathematics, University of Colorado, Boulder, where the work started initially. The authors thank Dr.~Judith Bunder in the Department of Mathematical Sciences, Adelaide University, for helping at various stages of the development of the numerical method.


\end{document}